\documentstyle[12pt]{article}

\newfont{\frak}{eufm10 scaled\magstep1}
\newfont{\extra}{msbm10 scaled\magstep1}
\newcommand{\fra}[1]{\mbox{\frak #1}}
\newcommand{\extr}[1]{\mbox{\extra #1}}

\newcommand{\sect}[1]{\setcounter{equation}{0}\section{#1}}
\newcommand{\subsect}[1]{\subsection{#1}}

\def\bea{\begin{eqnarray}}
\def\eea{\end{eqnarray}}
\def\beq{\begin{equation}}
\def\eeq{\end{equation}}
\def\ba{\begin{array}}
\def\ea{\end{array}}
\def\C{{\extr C}}
\def\R{{\extr R}}

\begin{document}

\begin{center}
{\large{\bf{DEFORMATION IN PHASE SPACE}}}
\end{center}

\vskip 1cm

\begin{center} 
Oscar Arratia $^{1}$, Miguel A. Mart¡n $^{1}$ and 
Mariano A. del Olmo $^{2}$
\end{center}

\begin{center}
{\it $^{1}$ Departamento de Matem tica Aplicada a la
Ingenier¡a,  \\
Universidad de Valladolid. E-47011, Valladolid, Spain.\\
E. mail: oscarr$@$wmatem.eis.uva.es, migmar$@$wmatem.eis.uva.es}
\end{center}

\begin{center}
{\it $^{2}$ Departamento de F¡sica Te¢rica,\\
Universidad de Valladolid. E-47011, Valladolid, Spain.\\ 
E. mail: olmo$@$cpd.uva.es}
\end{center}

\vskip 1cm
\begin{center}
\today
\end{center}

\begin{abstract}
We  review several procedures of quantization formulated in the
framework of (classical) phase space $M$. These quantization
methods consider Quantum Mechanics as a ``deformation" of Classical
Mechanics by means of the ``transformation" of the commutative algebra 
${\cal C}^{\infty}(M)$ in a new non-commutative algebra 
${\cal C}^{\infty}(M)_{\hbar}$. These ideas lead in a natural way to 
Quantum Groups as deformation (or quantization, in a broad sense) of
Poisson--Lie groups, which is also analysed here. 
\end{abstract}

\sect{Introduction}

Knowledge and understanding of Nature is the object of Physics.
The approach to the {\em real world} is  usually made in successive 
steps, in such a way that the old theory is recovered from the new
one by dropping the new effects. In practice this process is
carried out by a kind of limit procedure as is illustrated in the
following concrete situations.

>From the beginning of Einstein's Theory of Relativity, it is well-known 
that non-relativistic Classical Mechanics can be seen as the limit of 
Relativistic Mechanics when the speed of  light goes to infinity.
On the other hand, the Galilei group is a contraction, in the sense
of In\"on\"u and Wigner \cite{IW53}, of the Poincar' group when the
contraction parameter $\varepsilon=1/c$ goes to zero (Galilei and
Poincar' groups are the kinematical groups of the non-relativistic 
Classical Mechanics and the Relativistic Mechanics, respectively).

Another interesting example of this process relating two physical
theories is given by Quantum and Classical Mechanics, the latter can
be  considered as the limit of the first one when Planck's constant,
$\hbar$,  goes to zero. In both examples, Relativistic Mechanics and
Quantum  Mechanics depend on a parameter whose limit ($c\to \infty,\
\hbar  \to 0$) leads to a different physical theory.  
  
Deformation can be considered as a kind of inverse procedure of the
contraction of Lie groups as well as of the above limits for physical 
theories. From this viewpoint, the Poincar' group is a deformation of the 
Galilei group, Relativity Theory is a deformation of the non-relativistic 
Classical Mechanics and Quantum Mechanics is a deformation of Classical 
Mechanics. 

As is well-known, physicists have been very interested in finding
procedures that allow to  obtain quantum systems from classical ones,
i.e., to solve the problem  of the quantization of classical systems, 
from the first days of Quantum Mechanics onward.
Nevertheless, the deep differences  between both (classical and quantum)
theories at the level of  mathematical formalism as well as physical
interpretation have made impossible up to now to solve this problem in
a complete and satisfactory way. 

Among all the quantization procedures, stand out those of canonical 
quantization, geometric quantization \cite{Kos70,Sou70,Kir76,Woo92}, and 
group quantization \cite{AA82}, which are related with the usual 
formalism of quantum theory, as well as Moyal quantization \cite{Moy49}, 
Berezin quantization \cite{Ber75}, $*$--product formalism \cite{BFF78} 
and  Fedosov quantization \cite{Fed94} associated with the phase space
framework. 

The idea that quantization is deeply related with deformation was 
introduced by Bayen {\sl et al.} in \cite{BFF78}. For these authors, 
Quantum Mechanics can be replaced by a deformation of Classical Mechanics 
describing quantum systems in terms of functions defined on their phase
spaces. This can be achieved introducing a non-commutative product
($*$--product)  of these functions that replaces the usual commutative
product  of functions. The mathematical tool for this quantization
theory is the  deformation of Lie algebras \`a la Gerstenhaber
\cite{Ger64}.  A particular case of this
kind of deformation of  Classical Mechanics is the theory of Moyal
\cite{Moy49}.

It is worthy to note that quantization in terms of $*$--products (or
deformation) plays with respect to the formalism of  Quantum
Mechanics in phase space framework a similar role to that  geometric
quantization plays with respect to the standard formalism of Quantum
Mechanics, i.e, in terms of Hilbert spaces, operators, etc.

On the other hand, this formalism is closely related with quantum
groups,  which are deformed (Hopf) algebras, in the sense of
Gerstenhaber, of universal enveloping Lie algebras for quantum algebras,
or deformation of Poisson-Lie structures for quantum groups.
 
In this work we review different procedures of quantization of
classical systems from the optics of the deformation theory.
Incidentally, all of them try to formulate Quantum Mechanics in terms of
the formalism of phase space. A second part of this paper shows how
these ideas can be used in the theory of quantum groups. Now the objects
to deform are a kind of Poisson structures over a Lie group
(Poisson--Lie groups), which play in some sense the role of phase
spaces, and the $*$--product procedure can be implemented in order to
quantize or deform these objects giving rise to one of the few 
procedures to get quantum groups.

The paper is organized as follows. Section 2 presents the Moyal
quantization theory. When the physical system under study has a symmetry
group  one can profit this fact in order to systematize Moyal's
quantization by means of the Stratonovich--Weyl correspondence, and
this is the subject of Section 3. Two interesting examples are
showed to illustrate how the theory works. In the following
section we present a short review about the $*$--product.
Last section is devoted to Quantum Groups. We also show the
procedure allowing to obtain  Quantum groups starting from
``classical" structures like Poisson--Lie groups by means of a
$*$--product that deforms these objects. As an example we quantize the
group $SL(2)$.

\sect{Moyal's quantization}
 
The kinematical description of classical physical systems can be
modeled using a symplectic manifold 
$(M, \omega)$. The closed two-form $\omega$ identifies (sections of)
the tangent and the cotangent bundle on $M$. The dynamical behaviour 
of the system is then controlled by a function $H$ defined on the
manifold through the  vector field associated by $\omega$ to its
differential. This is the arena of Classical Mechanics, and the object
described  by $(M,\omega, H)$ is called a Hamiltonian classical system. 

The physical description of the previous system in terms of states and
observables carries a certain mathematical ``duality'' implemented by $M$ 
and the set of (smooth) functions $ {\cal C}^{\infty}(M)$. In a more 
technical language, we can say that this duality is realized by the 
contravariant Gelfan'd--Naimark functor, which shows that no 
information is lost if we replace the manifold $M$ by the algebra ${\cal 
A}= {\cal C}^{\infty}(M)$. From this point of view {\em every} structure 
defined on $M$ has a {\em natural} analogous on $\cal A$, in particular 
$\omega$ is transferred to a Poisson bracket on $\cal A$. Therefore, we 
can make a good definition for our system using the triplet $({\cal
A},\{\cdot,  \cdot\}, H)$ and this formalism permits an immediate
generalization: to consider algebras such that the commutativity
assumption is relaxed.

This procedure fits nicely into the problem of quantization because 
it is precisely what we are looking for when we try to ``quantize'' 
a classical system. Obviously, the passage from a commutative algebra
$\cal A$ to  a non-commutative one can be done in a large variety of
ways. Usually, one considers a new algebra ${\cal A}_h$ depending on
one or more parameters, and imposes that for a value of the parameter,
say $h=0$,  the algebra reduces to the commutative one. A far-reaching
idea is to build up the new algebra over the underlying set of the
algebra
$\cal A$ by means of a new product denoted by $*_h$. This product
allows us to define a new deformed Poisson bracket 
$$
\{f, g\}_h= f * g - g *f,\quad f,g \in {\cal A},
$$ 
which is a Lie algebra deformation of the original one. We will
show this construction later. 
  
The aim of Moyal's formulation of Quantum Mechanics is to describe it 
as a statistical theory taking place on a classical phase space unlike
the standard formulation, which is developed by means of Hilbert space
methods. In this way Moyal obtained a theory conceptually more
transparent (for more details see \cite{Gad95} and references therein). 
 
Within this framework observables and states of a quantum system are
considered as (generalized) functions on a phase space $M$ isomorphic
to $\R^{2n}$ (again the algebra ${\cal A}= {\cal C}^\infty(M)$). The
expectation value of the observable $A$ in the state $\rho$ is given by 
$$
\langle A \rangle_\rho= \frac{\int_{M} A \rho}{\int_M \rho},
$$
just like in classical statistical mechanics.
 
Moyal's formulation unifies in a single theory two important 
constructions: Weyl mappings and Wigner functions. For that reason we
call this theory the  Moyal--Weyl--Wigner formulation. 

Let us see what is the role played by Moyal's work in the problem  of 
quantization. The simplest and most usual quantization procedure is 
canonical quantization (or principle of correspondence). This scheme 
works rather well for physical systems whose phase space is isomorphic
to $\R^{2n}$, and it uses the Hilbert space $L^2(\R^n)$ of square
integrable  functions on $\R^n$ with respect to the Lebesgue measure.
This method of  quantization takes advantage of Dirac's prescription in
order to associate  functions (classical observables) with operators
(quantum observables). Thus, to the position and momentum coordinates,
$q_i$ and $p_i$, it associates the operators $Q_i$ (multiplication by
$q_i$) and  $P_i= -i  \hbar \frac{\partial}{\partial q_i}$,
respectively. The operator linked to the function
$f(q_i, p_i)$ is obtained formally replacing the  classical coordinates
by their corresponding operators, which yields  $f(Q_i, P_i)$. However,
operators
$Q_i$ and $P_i$ do not commute, and  henceforth the expression $f(Q_i,
P_i)$ is meaningless unless we fix some  ordering. 

The mathematical meaning of Dirac's prescription is as follows: if the 
Poisson bracket of two canonical coordinates is  
$$ 
\{q_i, p_j\}= \delta_{ij}, 
$$
then the commutator for the corresponding operators is 
$$  
[Q_i, P_j]= i \hbar \delta_{ij} .
$$
So, there is a faithful representation of the Lie subalgebra of 
$({\cal C}^\infty(M),\ \{\cdot, \cdot\})$, generated by the local
coordinates $(q_i,  p_i,\ i=1,\dots, n)$, in the Hilbert space 
$L^2(\R^n)$. In other words, we have a  homomorphism between the Lie
algebras of classical and quantum observables  (Heisenberg Lie algebra).
This last interpretation leads to the general rule of canonical
quantization  
$$
\{\cdot, \cdot \} \longrightarrow \frac{1}{i\hbar} [\cdot, \cdot].
$$ 
 
It is worthy to note that there are many possibilities to extend Dirac's  
prescription to general functions according to the ordering we select on 
monomials in  $Q_i$ and $P_i$. The most usual ones are the normal 
ordering  ($q^m p^n \longrightarrow Q^m P^n$), the antinormal ordering
($q^m p^n  \longrightarrow P^nQ^m $), and the Weyl ordering or Weyl's
correspondence  rule: 
$$
q^m p^n \longrightarrow (Q^m P^n)_S=
\frac{n! m!}{(n+m)!} {\displaystyle \sum_i} P_i^{m,n}(Q^m P^n),
$$
where $P_i^{m,n}$ are permutations with repetition of $m$ operators 
$Q$ and $n$ operators $P$.. Last ordering is the most suitable for 
quantum formulation on phase space. Moreover it exhibits invariance
under the Galilei and the symplectic groups.

A variant of canonical quantization is given by the Weyl postulate, 
which associates functions with operators along 
$$
 e^{\frac{i}{\hbar}({\bf x}\cdot {\bf p}+{\bf y}\cdot {\bf q})} 
\longrightarrow e^{\frac{i}{\hbar}({\bf x}\cdot {\bf P}+{\bf
y}\cdot {\bf Q})},
$$ 
where ${\bf x}, {\bf y} \in \R^n$. Note that canonical quantization can
be interpreted as a representation of the Heisenberg algebra, and
Weyl's quantization corresponds to a unitary representation of the
Heisenberg group. 

Now, if $f$ is a regular function on $\R^{2n}$ 
such that the Fourier transform $\hat{f}$ exists, then
$$
f({\bf p},{\bf q})= \frac{1}{(2\pi \hbar)^n} \int_{\R^{2n}} d{\bf x} \,
d{\bf y} \hat{f}({\bf x},{\bf y}) 
\ e^{\frac{i}{\hbar} ({\bf x}\cdot {\bf p}+{\bf y}\cdot {\bf q})} . 
$$
This expression, together with Weyl's postulate, leads to the natural 
definition of the Weyl correspondence, which associates the operator $W_f$ 
with the function $f$ by means of  
$$
W_f = \frac{1}{(2\pi \hbar)^n} \int_{\R^{2n}} d{\bf x} \, d{\bf y} 
\hat{f}({\bf x},{\bf y})  \ e^{\frac{i}{\hbar} ({\bf x}\cdot 
{\bf P} +{\bf y}\cdot {\bf Q})}.  
$$ 
Mapping $W$ can be extended to generalized functions  (i.e., 
distributions) on $\R^{2n}$. 

On the other hand, the function $\rho({\bf p},{\bf q})$ linked with 
the state operator $\rho$ is given by
$$
\rho({\bf p},{\bf q})=\frac{1}{(2 \pi \hbar)^n} 
\int_{\R^{2n}} d{\bf x} \, d{\bf y}\,
e^{\frac{i}{\hbar} ({\bf x}\cdot {\bf p} + {\bf y}\cdot {\bf q})} \;
{\mbox Tr} [\rho\, e^{\frac{-i}{\hbar} ({\bf x}\cdot {\bf P}
+{\bf y}\cdot {\bf Q})}]. 
$$
When $\rho$ is a pure state, i.e. $\rho=|\psi \rangle \langle \psi|$,
last expression reduces to 
$$
\rho({\bf p},{\bf q})=  \int_{\R^{n}} d{\bf x}
\ e^{\frac{i}{\hbar} {\bf x}\cdot
{\bf p}} \,\psi^*({\bf q}+\frac{1}{2} {\bf x}) 
\,\psi({\bf q}-\frac{1}{2} {\bf x}),
$$
which coincides with the expression given by Wigner \cite{Wig32}. 
Function $\rho(p,q)$ is called  Wigner's function.

In fact, mappings provided by  Weyl's correspondence rule and 
Wigner's functions are inverse of each other. An elegant proof 
of this fact uses the Grossmann--Royer operators \cite{Gro76,Roy77} 
defined by
$$
[\Omega({\bf p},{\bf q})\psi]({\bf x}):= 2^n \ 
e^{\frac{2i}{\hbar}{\bf p}\cdot ({\bf x}-{\bf q})} 
\,\psi(2 {\bf q}-{\bf x}). 
$$
So, the Weyl mapping can be rewritten as
$$
W_f = \frac{1}{(2\pi \hbar)^n} \int_{\R^{2n}} d{\bf p} \, d{\bf q} \,
f({\bf p},{\bf q}) \, \Omega({\bf p},{\bf q}).
$$
Using the complete set $\{ |{\bf x} \rangle, {\bf x} \in \R^n\}$ of 
kets for ${\bf Q}$ we can define the trace of an operator $A$ by
$$
{\mbox Tr}\, A= \int_{\R^n} \langle {\bf x}| A | {\bf x} \rangle .
$$
Computing the trace of the  product of two Grossmann--Royer operators, 
which has a distributional meaning, we obtain 
$$
{\mbox Tr}[\Omega({\bf p},{\bf q}) \Omega({\bf p}',{\bf q}')]= 
(2 \pi \hbar)^n  \delta ({\bf p}-{\bf p}') \delta ({\bf q}-{\bf q}').
$$
Finally, given an operator $A$ acting on $L^2(\R^n)$, its associated
function on phase space is 
$$
W^{-1}(A)= {\mbox Tr}[\Omega({\bf p},{\bf q}) A].
$$

As we said before, Moyal's contribution consists in combining both 
the  Weyl mapping and Wigner functions to construct a new product,
associative but non-commutative, for functions on phase space through
the equation 
$$
f * g := W^{-1} (W_f W_g).
$$
Moyal's product is stated in such a way that the the following 
diagram is commutative 
$$
\begin{array}{ccc}
 (W_f, W_g) & \longrightarrow & W_f W_g \\
W^{-1}\downarrow\qquad  & &\qquad  \downarrow W^{-1} \\
(f,g)  & \longrightarrow & f * g
\end{array}
$$
i.e., the quantum information encoded in the non-commutative product
of operators (quantum observables) is transferred via $W$ to the space
of classical observables and stored in the $*$--product.
Note that $W$ is a linear continuous map, $ W:S'(\R^{2n})
\longrightarrow {\cal L}(S(\R^{n}) ,S'(\R^{n}))$. However, Moyal's
product is not defined for all pairs of elements in $S'(\R^{2n})$.
There exists a maximal closed subspace ${\cal M}(\R^{2n})$ of 
$S'(\R^{2n})$ where the Moyal product is well defined. This space has the
structure of an algebra with respect to the sum of functions, product by 
scalars and Moyal product. We have the following chain of inclusions 
$$
{\cal S}(\R^{2n}) \subset {\cal L}^2(\R^{2n}) \subset
{\cal M}(\R^{2n}) \subset {\cal S'}^2(\R^{2n}). 
$$

The $*$--product can be expressed through the integral formula
$$
f*g = \frac{1}{(2\pi\hbar)^{4n}}\int_{\R^{4n}} d{\bf v} \, d{\bf w} \, 
f({\bf v}) g({\bf w}) \ 
e^{\frac{i}{\hbar}({\bf u}^tJ{\bf v}+{\bf v}^tJ{\bf w}+{\bf
w}^tJ{\bf u})},
$$
where $J=  \left( \begin{array}{cc} 0 & I_n \\ -I_n & 0 \end{array}
\right)$, ${\bf u}=({\bf q},{\bf p})^t, {\bf v}=({\bf q}',{\bf p}')^t$ 
and ${\bf w}=({\bf q}'', {\bf p}'')^t$. As a direct consequence, 
$*$ is a non  local product, but it reduces to a local one in the limit $\hbar 
\rightarrow 0$.

Using the canonical Poisson bracket on $\R^{2n}$
$$
f \stackrel{\leftrightarrow}{P} g\equiv \{f, g\}= 
\frac{\partial f}{\partial q_i} \frac{\partial g}{\partial p_i} - 
\frac{\partial f}{\partial p_i} \frac{\partial g}{\partial q_i}
$$
it is also possible to write down the Moyal product in differential 
form by means of the exponential of $\stackrel{\leftrightarrow}{P}$ 
$$
 f*g = f e^{-i \frac{\hbar}{2} \stackrel{\leftrightarrow}{P} } g.
$$
Thus, Moyal's product can be characterized as a bilinear and associative 
mapping in the following way \cite{BFF78,Ta90}. Let us consider the
bidifferential operator
$$
\begin{array}{cccc}  
{\cal J}_{\{\cdot, \cdot \}}: & 
{\cal C}^\infty(\R^{2n}) \otimes {\cal C}^\infty(\R^{2n}) &
\longrightarrow &
{\cal C}^\infty(\R^{2n}) \otimes {\cal C}^\infty(\R^{2n}) \\
& f\otimes g & \mapsto & \{f, g\}
\end{array}
$$
and  the product in ${\cal C}^\infty(\R^{2n})$
written as 
$$
\begin{array}{cccc}  
m: & {\cal C}^\infty(\R^{2n}) \otimes {\cal C}^\infty(\R^{2n}) &
\longrightarrow & {\cal C}^\infty(\R^{2n})  \\ 
& f\otimes g & \mapsto & f g,
\end{array}
$$
then the Moyal product can be expressed as 
\begin{equation}\label{pepe}
 *= m \circ e^{-i\frac{\hbar}{2} {\cal J}_{\{\cdot,\cdot\}} }.
\end{equation}

A remarkable property of the Moyal product is its equivariance under
transformations belonging to the symplectic or the Galilei groups, i.e.,
$$
(f_1 * f_2)^g= f_1^g \ * f_2^g,
$$
where $g$ is a generic group element and 
$f^g({\bf u})= f(g^{-1}{\bf u})$.

The Moyal bracket is defined antisymmetrizing the Moyal product, i. e.,
$$
\{f, g\}_{M}= \frac{1}{-i\hbar}(f * g-g * f)
$$
or 
$$
 \{f, g\}_{M}= \frac{1}{-i\hbar} W^{-1}[W_f,W_g].
$$

This bracket (playing the role of commutator in standard formulation) 
allows us to determine the evolution of the observable $f$ by
$$
\{H,f\}_M=\frac{df}{dt},
$$
where $H$ is the Hamiltonian of the system. In this dynamical 
sense we can translate some concepts of the standard formulation of
Quantum Mechanics to the new  one. Thus, given an evolution operator
$U(t)$, it is possible to construct a  Moyal propagator by
$$
 \Xi({\bf p},{\bf q},t)= W^{-1}(U(t)).
$$
If $H({\bf p},{\bf q})$ is a classical time-independent Hamiltonian its 
Moyal propagator is 
\begin{equation}
  \Xi_H({\bf p},{\bf q},t)= W^{-1}(e^{-\frac{it}{\hbar} W_H }).
\label{bw}
\end{equation}
The Schr\"odinger equation, $i \hbar \frac{\partial U(t)}{\partial t}= H 
U(t)$, can be rewritten in terms of the propagator (\ref{bw}) as
$$
 i \hbar \frac{\partial}{\partial t} \Xi_{H}({\bf p},{\bf q},t)
= H * \Xi_H({\bf p},{\bf q},t).
$$

We can also define the spectral projection by the Fourier transform of
the Moyal propagators with respect to the variable $t$ 
$$
\Gamma_H({\bf p},{\bf q},E)= \frac{1}{2\pi \hbar} \int_{\R} dt \,
        \Xi_H({\bf p},{\bf q},t)\,
      e^{-\frac{i}{\hbar}t E }.
$$
The support on $E$ (energy) of the projection associated to $H$ is, in a
large variety of cases \cite{BFF78}, the spectrum of the operator  $W_H$.

\sect{Stratonovich--Weyl correspondence} 

Symmetry principles play a central role in the analysis of physical
systems. In modern physics it is customary, given a Lie group $G$, to
define its associated classical elementary systems as $G$-homogeneous
symplectic spaces where the group acts by symplectomorphisms. After
the celebrated theorem by Kostant--Kirillov--Souriau \cite{Kir76} these
elementary systems are diffeomorphic to some orbit in $\fra {g}^*$ (the 
dual of the Lie algebra $\fra {g}$ of the group $G$) under the 
coadjoint action. This fact has as an immediate consequence the complete 
classification of all elementary systems whose symmetry is determined by 
the Lie group $G$. In a similar way, quantum elementary systems for $G$ 
are introduced  as projective unitary irreducible representations
(PUIR) of $G$ \cite{Wig39}.  Picking up some ideas from geometric
quantization, the link between  classical and quantum systems is
provided by Kirillov's theorem at least  for nilpotent groups
\cite{Kir76}. 

The previous definition of a quantum system fits quite
well in conventional formulation of Quantum Mechanics but we are
interested in quantum systems from Moyal's point of view. Therefore, we
adopt, as definition for a Moyal quantum elementary system
\cite{CGV90}, the pair formed by a  coadjoint orbit and a $*$--product
in the space of smooth functions defined on the coadjoint orbit. 

Kirillov's theorem can be considered as a partial answer to the 
problem of  quantization. Information provided by this geometric quantization can
be  best used to quantize a physical system according to Moyal's theory.


\subsect{Stratonovich--Weyl kernels}
 
 The tool required for passing from geometric quantization to Moyal
quantization is known as  Stratonovich--Weyl kernel \cite{Str57}, whose
definition is as follows. Given a $G$--coadjoint orbit $\cal O$ and
its corresponding PUIR $U$ with support space the
Hilbert space $\cal H$, the Stratonovich--Weyl  (SW) kernel is an
operator valued mapping 
$\Omega:{\cal O} \longrightarrow  {\cal L}({\cal  H})$ that satisfies the
following axioms:   
\begin{enumerate} 
\item $\Omega$ is injective, 
\item $\Omega({\bf x})$ is  self-adjoint  
$\forall \bf x \in {\cal O}$, 
\item unit trace: ${\mbox Tr}\,
\Omega({\bf x})=1,   \quad \forall {\bf x} \in {\cal O}$,
\item covariance: 
\begin{equation}
U(g) \Omega({\bf x}) U(g^{-1})= \Omega(g\cdot {\bf x}), 
\quad \forall {\bf x} \in {\cal O},\, \forall g\in  G.
\label{covariance}
\end{equation}
\item traciality:
\begin{equation}
\int_{\cal O} d\mu({\bf x}) {\mbox Tr}
[\Omega({\bf y}) \Omega({\bf x})] \Omega({\bf x})=
\Omega({\bf y}). \quad \forall {\bf y} \in {\cal O}, 
\label{traciality}
\end{equation}
where $\mu$ is a $G$--invariant measure on $\cal O$.
\end{enumerate}
Fourth axiom is a natural requirement in view of the symmetry of 
the system. Traciality means that the quantity 
${\mbox Tr}[\Omega({\bf y}) \Omega({\bf x})]$ essentially works like 
Dirac's distribution $\delta({\bf y}-{\bf x})$.

The SW kernel allows us to build up a symbol calculus. The
``symbol'' associated with an operator $A$ is given by
\begin{equation}
{\cal W}_A({\bf x})= {\mbox Tr}[A \Omega({\bf x})], \label{cc}
\end{equation}
and  the mapping 
$$
\begin{array}{ccc}
        {\cal L(H)} & \longrightarrow & {\cal C}^\infty(O) \\
           A        &  \mapsto &  {\cal W}_A,  
\end{array}
$$
is called the SW correspondence. Observe that now the mapping 
${\cal W}$ is the inverse of the mapping $W$ defined in section 2. 
It is worthy to note that if $A \mapsto {\cal W}_A$ is injective then 
expression (\ref{cc}) can be inverted as
$$
A= \int_{\cal{O}} d\mu({\bf x}) \, {\cal W}_A({\bf x}) \Omega({\bf x}),
$$
which shows that the same kernel implements both directions $A  
\leftrightarrow {\cal W}_A$ of the correspondence. Sometimes it is said 
that $\Omega$ is a quantizer and also a ``dequantizer''.

The properties satisfied by the kernel $\Omega$ have immediate
consequences  on the SW correspondence, remarkable ones are: 
\begin{itemize}
\item Symbols associated to selfadjoint operators are real,
$$
A= A^\dagger\  \Rightarrow {\cal W}_A^*= {\cal W}_A. 
$$
\item The identity operator has as symbol the unit function,
$$  
{\cal W}_I= 1.
$$
\item Covariance condition leads to 
$$
{\cal W}_{U(g)AU(g^{-1})}(g\cdot {\bf x})= {\cal W}_A({\bf x}).
$$
\item The trace of the product of two operators can be evaluated as 
an integral involving their symbols
$$
{\mbox Tr}(AB)= \int_O d\mu({\bf x}) \, {\cal W}_A({\bf x}) 
{\cal W}_B({\bf x}). 
$$
\end{itemize}
Another crucial application of the SW kernel is the construction of a
non-commutative (or twisted)  product 
$$
(f * g)({\bf x})= \int_{\cal O} d\mu({\bf y})  \,  \int_{\cal O}
d\mu({\bf z})\,  L({\bf x},{\bf y},{\bf z}) 
f({\bf y}) g({\bf z}),
$$
where $L({\bf x},{\bf y},{\bf z})= {\mbox Tr}[\Omega({\bf x})
\Omega({\bf y}) \Omega({\bf z})]$, and is called trikernel. 
This construction of the $*$--product
assures that the SW correspondence is an algebra morphism
$$
{\cal W}_{AB} = {\cal W}_A * {\cal W}_B.
$$
Another interesting equality involving averages is 
$$
\int_{\cal O} d\mu({\bf x}) (f_1*f_2)({\bf x}) = \int_{\cal O} 
d\mu({\bf x}) f_1({\bf x}) f_2({\bf x}).
$$
The geometrical meaning of covariance is reflected in the 
$G$--equivariance of the $*$--product
$$
(f_1*f_2)^g= f_1^g * f_2^g, \qquad \forall g \in G.
$$
At the level of the trikernel it means invariance, i.e.,
$L(g\cdot {\bf x}, g\cdot {\bf y}, g\cdot {\bf z})= 
L({\bf x},{\bf y},{\bf z})$.

Up to now, no general result guaranteeing the existence of $\Omega$ is 
available. A practical recipe to build a SW kernel is summarized in 
the three following steps: 
\begin{enumerate}
\item Select a point ${\bf 0}$ as ``origin" of the orbit and take a 
section  $s: {\cal O} \to G$, ($\cal O$ is viewed as the homogeneous
space
$G/G_0$, where  $G_0$ is the isotopy group of the origin).
\item Choose an operator $A$ as Ansatz for the value of $\Omega$
at the  origin ($A=\Omega(0)$). If $A$ is a good
Ansatz it suffices with fixing the  kernel on the whole orbit because
of covariance (\ref{covariance}) one gets 
$$
\begin{array}{rl}
\Omega({\bf x})&= \Omega(s({\bf x})\cdot {\bf 0})\\[.3cm]
&=  U(s({\bf x})) \Omega({\bf 0}) U(s({\bf x})^{-1})= 
U(s({\bf x})) A U(s({\bf x})^{-1}).
\end{array}
$$
\item Verify that $A$ is indeed a good Ansatz checking  the
mapping that it determines satisfies all the axioms needed.
\end{enumerate}
Obviously, the first axiom to be checked is covariance, if it 
fails $\Omega$ can not be defined!. For this purpose it is useful
the following lemma \cite{Arr96,Mar98}.

\medskip
{\bf Lemma 3.1}. {\sl  Propositions (a), (b) and (c) are equivalent}
\begin{description}
\item[(a)]  $\Omega({\bf x})= U(g) \Omega({\bf x}) U(g^{-1}), \quad
\forall  ({\bf x},g)\in ({\cal O}, G)$,\\ i.e. $\Omega$ {\sl verifies
the covariance axiom (\ref{covariance})},
\item[(b)]  $\Omega({\bf 0})= U(g) \Omega({\bf 0}) U(g^{-1}), 
\quad \forall g\in G_0,$
\item[(c)]  $[U(X), \Omega({\bf 0})]= 0, \quad \forall X \in
{\frak g}_0$,\\ where ${\fra g}_0$ is the Lie subalgebra
associated with the isotopy subgroup.
\end{description}

The choice for the Ansatz $A$ is heavily based on parity-like 
operators due to the form of the Grossmann--Roger operator. 
Nevertheless, this choice does not successfully leads to a SW kernel in many cases. 

In the following we present two examples that illustrate how the theory
works (for more details see \cite{Arr96,Mar98,Arr97} and references
therein).

\subsect{Example 1: Galilean systems in $(1+1)$ dimensions}

The Galilei group is the set of transformations that relate 
observables measured from different inertial frames in non-relativistic
mechanics. The Galilei group  can also be defined from an active point 
of view in which galilean transformations (time and space
translations, galilean boosts and space rotations) act on the
space-time manifold. In $(1+1)$ dimensions that  action is given by
$$
  (t',x')\equiv (b,a,v)\cdot (t,x)= (t+b, x+a+vt),
$$
where $b, a$ and $v$ denote the parameters of time and space
translations and galilean boosts, respectively. The transformations
$(b,a,v)$ form a Lie group denoted $G(1,1)$, whose composition  law is
obtained from the previous action 
$$
  (b',a',v')(b,a,v)= (b'+b, a'+a+v'b, v'+v).
$$
Its associated Lie algebra  ${\fra g}(1,1)$ is
spanned by the infinitesimal generators of time ($H$) and space ($P$)
translations and galilean boosts ($K$), which have the following
commutation relations  
$$
  [K,H]=P, \quad [K,P]=0, \quad [P,H]=0.
$$
Elements of the dual space ${\fra g}^*(1,1)$ of the Lie algebra 
${\fra g}^(1,1)$ are linear
combinations,  $h H^* + p P^* + k K^*$, in terms of the dual basis of
$\{H,P,K\}$. The   coadjoint action of $G(1,1)$ on ${\fra g}^*$ is
expressed in that  coordinates  as 
$$ 
 (h',p',k')\equiv (b,a,v)\cdot (h,p,k)= (h-vp, p, bp+k). 
$$ 
The space ${\fra g}^*(1,1)$ is then ``foliated'' into two kind of
orbits:
\begin{enumerate}
\item 0--dimensional (0D) orbits : points of the form $h H^*+ k K^*$, 
\item 2--dimensional (2D) orbits ${\cal O}_\alpha$: characterized by
equation  $p=\alpha$. 
\end{enumerate}
>From a physical point of view 0D orbits cannot support a 
dynamics, and hence they are not interesting. 
In the orbit ${\cal O}_\alpha$ 
a set of canonical coordinates is determined by $q=\frac{1}{\alpha} k$
and 
$p=h$. Taking ${\bf 0}= \alpha P^*$ (with canonical coordinates 
$(0,0)$) as  representative point on ${\cal O}_{\alpha}$ we can find a
maximal subordinate subalgebra, which induces by Kirillov's method 
the PUIR of $G(1,1)$ 
$$
 [U_\alpha(b,a,v) \psi](w) = e^{-i\alpha(a-bw)}\psi(w-v),    
$$
realized on the Hilbert space $L^2(\R)$ of square integrable functions on 
the real line. The argument of those functions can be identified with  
velocity. 

\medskip 
Let us construct a SW kernel following the three steps mentioned above: 
\begin{enumerate}
\item We had already chosen the origin ${\bf 0}= \alpha P^* \equiv (0,0)$. 
A normalized  section 
      is given by $s(p,q)= (q, 0, -\frac{p}{\alpha})\in G(1,1)$.
\item As Ansatz for the kernel at the origin we take the parity-like 
operator
$$
      [Af](\omega)= 2 \psi(-\omega).
$$
\item It is easy matter to check all the axioms. For instance, 
``unit trace'':
$$\begin{array}{rl}
     {\mbox Tr} \, \Omega(p,q)&= 
     {\mbox Tr}[U(s(p,q)) \Omega(0,0) U(s(p,q)^{-1})]=
     {\mbox Tr}[\Omega(0,0)]\\[.3cm]
&=\int_{-\infty}^\infty dw \, \langle w | \Omega(0,0) | w \rangle= 
       \int_{-\infty}^\infty dw \, 2 \langle w  |- w \rangle\\[.3cm]
&= \int_{-\infty}^\infty dw \, 2 \delta(2w)= 1. 
      \end{array}
$$
To prove covariance (\ref{covariance}) it suffices to consider the
isotopy group  of ${\bf 0}$ which is made up of spatial
translations 
$G(1,1)_{\bf 0}=\{(0,a,0)\}$. 
\end{enumerate} 

Let us show that, in this case, there exist many different SW
kernels.  Firstly,  note that covariance (i.e., 
$U_\alpha(0,a,0) A U_\alpha(0,-a,0)= A$) is satisfied
for any operator $A$ because the elements of the isotopy group are
represented by scalars. 

To analyse traciality (\ref{traciality}) we write the action of
$A$ as
$$
    [A \psi](w)= \int_{-\infty}^{\infty} dw'\, A_{w,w'} \psi(w'),
$$ 
this leads to the rather complicated condition
$$
 A_{w',w}= 2 \pi  \int_{-\infty}^{\infty} dv 
\int_{-\infty}^{\infty} dp\,
      A_{v+w'-w, v}\; \overline{A}_{v+w'-w+\frac{p}{\alpha}, v+ 
\frac{p}{\alpha}}\; 
      A_{w'+\frac{p}{\alpha}, w+\frac{p}{\alpha}},
$$
where the bar stands for complex conjugation. However, a particular
solution can be found
$$
    A_{w',w}= e^{i\varphi(w)} \delta(w+w'),
$$
and the associated kernel acts as
$$
  [\Omega(p,q)\psi](w)= e^{i\varphi(w +\frac{p}{\alpha})}
              e^{2i \alpha q (w+ \frac{p}{\alpha})} 
\psi(-w- 2\frac{p} {\alpha}).
$$
Finally to satisfy ``hermiticity'' and ``unit trace'' it is enough that 
$\varphi$ verifies $\varphi(w)+ \varphi(-w) \in 2 \pi {\extr Z}$ and  
$\varphi(0) \in 2 \pi {\extr Z}$, respectively.

\subsect{Example 2: The Newton--Hooke group $NH(1,1)$}

The kinematical group of 
Newton--Hooke in $(1+1)$ dimensions can be defined, 
in analogy with the case of the Galilei group, as the set of 
transformations (time and space translations and boosts) 
acting on the space-time as 
$$
  (t',x')\equiv (b,a,v)\cdot (t,x)= (t+b, x + a \cos\frac{t}{\tau} 
+ v \tau \sin  \frac{t}{\tau}).
$$
The natural topology of this universe is that 
of the product $\R\times S^1$. The parameter $\tau$,
with  dimension of time,  characterizes the compact direction, and can
be seen as a characteristic time of this universe. In the  definition of
the action we have opted for the universal covering to obtain  simpler
formulas (i.e., $b\in \R$). The group law is 
$$
  (b',a',v')(b,a,v)= (b'+b, a' \cos \frac{b}{\tau} + 
v' \tau \sin \frac{b}{\tau} 
+a, 
     v' \cos \frac{b}{\tau} - \frac{a'}{\tau} \sin \frac{b}{\tau} + v).
$$
Its associated Lie algebra, ${\fra nh}(1,1)$, spanned by the
infinitesimal generators of the above mentioned transformations, $H,\ P,\
K$, has only two non-vanishing commutators 
$$
  [K,H]= P, \quad [P, H]= -\frac{1}{\tau^2} K.
$$
The coadjoint action of $NH(1,1)$ 
$$
 (b,a,v) \cdot (h,p,k)= (h-vp+\frac{1}{\tau^2} a k,
    p \cos \frac{b}{\tau} - \frac{k}{\tau} \sin \frac{b}{\tau}, 
    p \tau \sin \frac{b}{\tau} + k \cos \frac{b}{\tau}),
$$
splits the dual 
${\fra nh}^*(1,1)$ 
into 0D and 2D orbits, points of the form $hH^*$, and cylinders 
${\cal O}_\beta$ defined
by $p^2 +  \frak{k^2}{\tau^2}= \beta$, respectively. 
 
A local chart of canonical coordinates is formed by the pair $\alpha= 
\arctan \frac{k}{\tau p}$, $j=\tau h$. For later use we quote
 here a normalized section
$$
  s(\alpha, j) = (\tau \alpha, 0,0) (0,0,-\frak{j}{\beta \tau}).
$$ 

In order to apply Kirillov's method for induced representations we take
the subalgebra  $\langle P,K \rangle$ which is subordinated to the point
$\beta P^* \in  {\cal O}_\beta.$
The PUIR attached to ${\cal O}_\beta$ can then  be realized on the space
$L^2([-\pi,\pi])$ of square integrable functions on the  circle
$$
 [U_\beta(b,a,v)\psi](t)=  e^{i \beta (\frac{a}{\tau}\cos t
+ v \sin t)}\psi(t-b). 
$$
Note that all the orbits ${\cal O}_\beta$ are diffeomorphic and all
the PUIR's $U_\beta$ have the same form, hence we will take
$\beta=1$.

Parity-like operators that the following ones
$$
    \mbox{[}A \psi \mbox{]}(t)= 2 \psi(-t), \quad
    \mbox{[}A \psi \mbox{]}(t)= 2 \psi(t+2\pi), \quad
    \mbox{[}A \psi \mbox{]}(t)= 2 \psi(-t+\pi), 
$$
as Ansatzs for $A$ do not yield SW kernels, the first one is not
tracial and the others  do not  accomplish covariance
(\ref{covariance}).  To solve this problem we start with a general
operator  and then impose consecutively all the axioms.  An operator
$A$ acting on
$L^2([-\pi,\pi])$  can be  expressed as 
$$
    A= \sum_{r,s\in {\extr Z}} A_{r,s} |r\rangle \langle s|,
$$
where the ket $|r\rangle$ stands for the function 
$\psi_n(t)=\frac{1}{\sqrt{2\pi}} e^{int}$. To apply covariance let
us note  that the isotopy group of the origin $o= \beta P^*$
contains only space  translations. Therefore, according Lemma 3.1
the covariance condition  reduces to $[U(P),A]=0$, whose most
general solution reads
$$
  [A\psi](t)= a(t)\psi(-t)+ b(t) \psi(t),
$$ 
with $a$ and $b$ arbitrary functions on $[-\pi, \pi]$.
Hence, the most general mapping $\Omega: {\cal O} \rightarrow {\cal 
L}(L^2([-\pi,\pi]))$,
verifying covariance is
$$
  [\Omega(j, \alpha)\psi](t)= e^{2ij \sin (t-\alpha)} a(t-\alpha) 
\psi(2 \alpha -t) + b(t-\alpha) \psi(t).
$$

The following lemma is useful to improve traciality.

{\bf Lemma 3.2:} 
{\sl A covariant mapping $\Omega: {\cal O} \rightarrow {\cal 
L}(L^2([-\pi,\pi]))$ verifies traciality if and only if $K(x,y)= {\mbox 
Tr}[\Omega(x) 
\Omega(y)]$ is a reproducing kernel in the space of symbols generated by 
$\Omega$, i.e.}, 
$$
    \int_{\cal O} d\mu(y) K(x,y) {\cal W}(y)= {\cal W}(y), 
\quad \forall x \in  {\cal O}.
$$
The proof of this lemma involves only covariance of $\Omega$ and
invariance  of  the measure $\mu$ under $NH(1+1)$. In fact, last
condition is equivalent to the following  one  
$$
    \int_{\cal O} d\mu(y) K(o,y) {\cal W}(y)= {\cal W}(y),
$$
apparently weaker.

Imposing traciality and the other axioms we obtain the
following  family of SW kernels on the cylinder
$$
  [\Omega(j, \alpha)\psi](t)= e^{2ij \sin (t-\alpha)} a(t-\alpha) 
\psi(2 \alpha -t), 
$$
where function $a$ is essentially arbitrary, only subject to the 
constraints $a(-t)= \overline{a(t)}$ and $|a(t)|^2+ |a(t+\pi)|^2
= 4 |\cos t|$. 

This solution  solves the Moyal quantization of the cylinder. 

\sect{Star products}

The theory of deformations of algebras of classical observables, 
called the theory of $*$--products, was introduced by  Bayen 
{\sl et al.} \cite{BFF78} and its mathematical foundations can be
found in the  works of Gerstenhaber \cite{Ger64} about deformation of
algebraic structures. As we said before in section 2 Moyal's
quantization can be  seen as a particular case of this theory.

Let ${\cal A}$ be an algebra and ${\cal A}[[h]]$ the algebra of formal 
power series in $h$ with coefficients in ${\cal A}$. The algebra 
${\cal A}[[h]]$ is said to be a deformation of ${\cal A}$ with 
deformation parameter $h$ if  
$$
{\cal A}[[h]]/h{\cal A}[[h]] \simeq {\cal A}.
$$

Let $(M,\ \{\cdot, \cdot\})$ be a Poisson manifold, and let us 
consider the Lie algebra  ${\cal A}={\cal C}^\infty (M)$. A quantization
of ${\cal A}$ is a deformation of the commutative algebra ${\cal A}$ into a
non-commutative algebra ${\cal A}_h= {\cal A}[[h]]$ with a new product, 
$*_h: {\cal A}_h \times {\cal A}_h \longrightarrow {\cal A}_h$, 
defined as a deformation of the commutative product on  ${\cal A}$. 
Since the elements of ${\cal A}_h$ are formal series
$$
f=f(x,h)=\displaystyle{ \sum_{r=1}^\infty} f_r(x) h^r, \qquad
f_r \in  {\cal C}^\infty (M),\ \ x \in M,  
$$
the $*$--product is defined as
$$
f *_h g =\displaystyle {\sum_{r=1}^\infty} l_r(x) h^r,
$$
such that $l_r$ are polynomials on $f_r$, $g_r$ and their 
derivatives, and $l_0(x)=f_0(x)g_0(x)$. Moreover, the $*$--product
should be associative. 

A commutator is defined on ${\cal A}_h$ by
$$
[f , g] \equiv \{f,g\}_h = f*_h g- g*_h f =h\{f_0,
g_0\} + o(h^2).
$$
Consequently 
\begin{equation}
f *_h g = fg+ \frac{h}{2} \{f, g\} + o(h^2),
\qquad \forall f,g\in  {\cal A} \label{cf}
\end{equation}
and 
\begin{equation} 
f *_h a = a *_h f = af, \qquad \forall a\in \C, \ \ f\in {\cal A}.
\label{cg}
\end{equation}
In particular $f *_h 1 = 1 *_h f = f$, i.e., the unit element
is not quantized.

Note that the original Poisson bracket on ${\cal A}$ is recovered in the
semiclassical limit ($h \to 0$)
$$
\{f, g\}= \lim_{h\rightarrow 0} \frac{1}{h} \{f,g \}_h.
$$

Additional conditions have to be added in physical situations:
$\overline{f *_h g}  = \overline{f} *_h \overline{g}$, and $h= -i
\hbar$. Therefore, after quantization real-valued classical
observables go  over into self-adjoint quantum observables
(operators).

In general a $*_h$--product is defined as
$$
f*_h g = fg + {\displaystyle \sum_{r=1}^\infty} C_r(f,g) h^r,
$$
where the terms $C_r$ are Hochschild 2-cochains, i.e.,  
bidifferential operators on ${\cal A}$ without constant term in each 
argument ($C_r$  vanishes over constants).

If the infinite series 
${\displaystyle \sum_{r=1}^\infty } C_r(f,g) h^r$ stops at order $m$
verifying the associativity property up to this order,  we have a
deformed product up to order $m$. Gerstenhaber showed that the
obstruction to the extension of deformed products up to order $m$
is the third space of the Hochschild cohomology.

There are some results about the existence of $*$--products based on
cohomological techniques involving the Hochschild or de Rham
cohomologies. The most interesting of them proves the existence of a
$*$--product for any symplectic manifold \cite{DL83}. Unfortunately, 
all these results are formal in the sense that  do not give an
effective or canonical construction procedure of such $*$--products.
As far as we know the most interesting example  of $*$--product
quantization is the Moyal--Weyl--Wigner quantization. 

It is worthy to note that, from a physical point of view, in this
quantization method the 
$*$--products have to be invariant for the elements (distinguished
observables) of a sufficiently large finite  subalgebra  ${\cal I}$
of ${\cal A}$ (i.e., $\{a,f_1*f_2\}=\{a,f_1\}*f_2+f_1*\{a,f_2\},\
\ a \in {\cal I},\ \forall f_1,f_2 \in {\cal A}$) such that
$[a,f]=h\{ a,f\}, \forall a\in {\cal I}, \ \forall f \in {\cal A}$. 
These distiguished observables determine a (local) coordinate
system of $M$ in terms of a basis of this subalgebra.  For instance,
in the case of  Moyal's product for $M=\R^{2n}$ the polynomials of
degree lesser or equal to two of the usual coordinates  ($q_i,
p_i,\ i=1,\dots, n$) constitute this subalgebra of distinguished
observables. The fact that the  quadratic Hamiltonians belong to
this subalgebra makes easier the study of the temporal evolution of
the quantized systems in this phase space  framework.  
 
\sect{Quantization of Poisson--Lie structures}

Quantum groups are objects which can be seen as deformation (or
quantization in a broad sense) of classical structures related with
${\cal C}^{\infty}(G)$, where $G$ is a Lie group (see \cite{CP} for a
review). 

The theory of $*$--products is one of the different approaches to
quantum groups (the others are FRT method \cite{FRT}, matrix $T$ 
\cite{FG}). It tries to construct $*$--products on 
${\cal C}^{\infty}(G)$ that quantize this algebra and preserve in some
sense its additional algebraic structure \cite{Ta90,DR}. 

The aim of this section is to quantize (or deform)
the corresponding Poisson algebra of classical observables ${\cal A}=
{\cal C}^\infty(G)$ of smooth functions over a Lie group $G$, which has 
a supplementary algebraic structure of Hopf coalgebra. We will start by
a brief review about these structures. 

\subsect{Poisson--Lie groups}

The Hopf coalgebra structure on  ${\cal A}= {\cal C}^\infty(G)$ 
(also denoted $Fun(G)$) is induced by the  composition law of the Lie
group
$G$. Explicitly, there are two homomorphisms: coproduct  ($\Delta:{\cal
A}\otimes {\cal A} 
\to {\cal A}$) and counit ($\varepsilon :{\cal A}\to \C$), and the
antihomomorphism antipode ($\gamma :{\cal A}\to {\cal A}$) defined by   
\begin{equation}
(\Delta f)(g, g')= f(gg'), \qquad
\varepsilon (f)= f (e), \qquad [\gamma(f)](g)= f(g^{-1}),
\label{coprod}
\end{equation}
$\forall g, g'\in G$, with $e$ the unit element of $G$. So,  
${\cal A}$ is said to be a Hopf algebra.

A Lie group $G$ is  a Poisson--Lie (PL) group if  both structures,
Poisson manifold and Hopf coalgebra, are compatible in the following
way:

1) Coproduct verifies the following diagram
$$
\begin{array}{ccc}
& {\{\cdot, \cdot\}}& \\
{\cal A}\otimes {\cal A} & \longrightarrow & {\cal A} \\
\Delta\otimes \Delta \downarrow \hspace{3em} &  & 
 \hspace{1em} \downarrow \Delta \\
& {\{\cdot, \cdot\}_{{\cal A}\otimes {\cal A}}} & \\
{\cal A}\otimes {\cal A} \otimes {\cal A}
 \otimes {\cal A} & \longrightarrow & {\cal A}\otimes
A,  \end{array}
$$
or in other words,
$$
\Delta(\{f_1, f_2\})= \{\Delta(f_1), 
\Delta(f_2)\}_{{\cal A}\otimes {\cal A}},
$$
where 
$$
\{ f_1 \otimes f_2, k_1 \otimes k_2 \}_{{\cal A}\otimes {\cal A}}= 
f_1 k_1 \otimes \{f_2, k_2\} + \{f_1, k_1\} \otimes f_2 k_2.
$$

2) The group multiplication law, $m: G \times G \longrightarrow G$, 
is a Poisson map.

An interesting problem is to construct 
PL structures over a Lie group. A solution is as follows. Let
us consider $r \in \Lambda^{2} {\fra g}$ (${\fra g}\ $= Lie($G$)), i.e., 
$r=r^{ij}X_i \otimes  X_j$ with $r^{ij}=-r^{ji}$ in a basis $\{X_i\}$ of
$\fra g$. If we define  
$$
r_{12}= r^{i,j} X_i \otimes X_j \otimes 1, \quad 
r_{13}= r^{i,j} X_i \otimes 1 \otimes X_j, \quad
r_{23}= r^{i,j} 1 \otimes X_i \otimes X_j, \quad
$$ 
the Schouten bracket of $r$ with itself can be written as 
$$
 [[r,r]]= [r_{12}, r_{13}] + [r_{12}, r_{23}] +[r_{13}, r_{23}].
$$

We say that $r$ is a classical $r$--matrix verifying the classical 
Yang--Baxter equation (CYBE) (or a  nonstandard $r$--matrix) if it
verifies $[[r, r]]= 0$.
When $[[r, r]]\neq 0$ but ${\mbox  {\rm ad}}_{\fra g}^{\otimes
3}[[r,r]]=0$, it is said that 
$r$ satisfies the modified classical Yang--Baxter equation (MCYBE)  or
is a standard $r$--matrix.

On the other hand, to every $X\in {\fra g}$ there are
left-invariant and right-invariant vector fields defined by
\begin{equation}
(X^L f)(g)= \frac{d}{dt}\Bigg\vert_{t=0} f(ge^{tX}), \qquad
(X^R f)(g)= \frac{d}{dt}\Bigg\vert_{t=0} f(e^{tX} g). \label{vecfields}
\end{equation}

Using both ingredients \cite{Ta90,DR}, a classical $r$--matrix and 
invariant vector fields, it is possible to endow $G$ with a structure
of  PL group  $(G, \{\cdot, \cdot\})$ with the Poisson bracket
defined by  (Sklyanin bracket) 
\begin{equation}
\{f_1, f_2\}=  r^{ij}(X_i^Rf_1 X_j^Rf_2- X_i^Lf_1 X_j^Lf_2), 
\quad f_1,f_2 \in {\cal C}_{\infty}(G). \label{sb} 
\end{equation}

We have seen that a PL group is a Poisson--Hopf algebra, 
hence it is natural to look for a Hopf algebra structure
on the deformation ${\cal A}_h$, or in other words, if it exists a
coproduct $\Delta_h$  (besides a counit and an antipode) such that
$$
\Delta_h(f_1 *_h f_2)= \Delta_h(f_1) *_h \Delta_h(f_2),
$$
and obviously $\lim_{h\rightarrow 0} \Delta_h= \Delta_{\cal A}$. The
noncommutative Hopf algebra obtained in this way will be called the
quantum group associated to $G$ (usually denoted $Fun_h(G)$).

\subsect{Quantization of PL groups with CYBE $r$--matrix}

The procedure of quantization of PL groups  presents some differences 
according with the associated $r$--matrix be standard or non-standard.
The easier case corresponds to non-standard $r$--matrices,
consequently  we will present firstly this case (for more details  
see \cite{Ta90,DR}). 

Although the Poisson brackets 
$\{f_1, f_2 \}_{L,R}= r^{ij} X_i^{L,R} f_1 X_j^{L,R} f_2$ do not
generate separately a PL structure on G (see expression (\ref{sb})),
the idea is to quantize separately each  Poisson structure
determined by each of the Poisson brackets and then to join them to
get a quantization of the PL group.

For instance, let us consider the left Poisson bracket, and let us go 
to find a left $G$--equivariant associative $*$--product
verifying conditions (\ref{cf}) and (\ref{cg}), and moreover
\begin{equation}
 \Delta(f_1 *_h f_2)= \Delta f_1 *_h \Delta f_2,\label{deltacoprod}
\end{equation}
where $ \Delta$ is the standard coproduct on ${\cal A}$ defined by
(\ref{coprod}a). The expression for the $*$--Moyal's product (\ref{pepe}),
$ *_{\mbox Moyal}= m \circ e^{\frac{h}{2} {\cal J}_{\{ .,.\}} }$,
suggests to define $*_h^L$ in the same way. So, 
$$
*_h^L= m \circ \tilde{F}, 
$$
where $\tilde{F}$ is a formal power series in $h$ whose coefficients 
 $\tilde{F}_n$ are linear differential operators on 
${\cal A}\otimes {\cal A}$,
i.e., 
$$
   \tilde{F}= 1 + \sum_{n=1}^\infty h^n \tilde{F}_n, \qquad
   \tilde{F}_n: {\cal A} \otimes {\cal A} \longrightarrow 
{\cal A} \otimes {\cal A}.
$$
The property of left-invariance for $*_h^L$ gives rise to
$$
(L_{g_1} \otimes L_{g_2}) \circ \tilde{F}_n=
  \tilde{F}_n \circ  (L_{g_1} \otimes L_{g_2}), \quad
   \forall g_1, g_2 \in G,
$$
where $L_g$ is the left-translation operator.

Let $\pi_L$ be the representation of the universal enveloping algebra
$U{\fra g}$ by left-invariant differential operators on 
${\cal C}^\infty(G)$ such that in terms of a basis $\{X_i\}$ of ${\fra
g}$ takes the form $\pi_L(X_i)= X_i^L$,  then $\tilde{F}_n$ can be
expressed as 
$$
\tilde{F}_n= (\pi_L \otimes \pi_L)(F_n), 
$$
with $F_n\in U{\fra g} \otimes U{\fra g}$ and $F_1= -\frac{1}{2} r$.
Hence $\tilde{F}$ can be written as the image by the representation
$\pi_L$ of a formal power series in $h$ with coefficients in 
$U{\fra g} \otimes U{\fra g}$:
$$
 \tilde{F}= (\pi_L \otimes \pi_L)(F), \qquad F\in 
 U{\fra g} \otimes  U{\fra g} [[h]].
$$
In order to get a deformation with non quantized unit and associative 
it is necessary to impose the following conditions on $F$:
\begin{equation}
\begin{array}{c}
  (\epsilon \otimes {\mbox id})F= ({\mbox id} \otimes \epsilon) F= 1,
\\
  (F \otimes {\mbox id})(\Delta_0 \otimes {\mbox id}) F=
  ({\mbox id} \otimes F)({\mbox id} \otimes \Delta_0)F,
\end{array}\label{conditionsF}
\end{equation}
where $\Delta_0$ denotes the coproduct in the Hopf
algebra $U{\fra g}$, i.e.,
$$
   \Delta_0(X) = 1 \otimes X + X \otimes 1, \qquad X \in {\fra g}.
$$
It can be shown that the $*$--product given by 
$$
  *_h^L= m\circ (\pi_L \otimes \pi_L)(F)
$$
defines  a left-invariant quantization of the Poisson bracket 
$\{., .\}_L$.

Similarly, the right-invariant $*$--product 
$$
  *_h^R= m\circ (\pi_R \otimes \pi_R)(F^{-1})
$$
quantizes the Poisson bracket $\{\cdot, \cdot\}_R$.

Finally, the combination of both $*_h^L$ and
$*_h^R$  
\begin{equation}
*_h= m \circ (\pi_L \otimes \pi_L)(F) \circ (\pi_R \otimes \pi_R)
     (F^{-1}) \label{starprod}
\end{equation}
yields an associative quantization of the PL group verifying
(\ref{deltacoprod}). 

The existence of an element $F\in U{\fra g} \otimes U{\fra g}[[h]]$ 
verifying (\ref{conditionsF}) for a nonstandard $r$--matrix of 
$\fra g$ has been proved by Drinfel'd \cite{DR}. This result assures
that any PL group associated to a  classical $r$--matrix verifying CYBE
can be quantized.

Introducing the flip operator $\sigma(a \otimes b)= b \otimes a$ it is
possible to construct an object, 
\begin{equation}
{\cal R}= \sigma(F^{-1}) F= 1- h r + o(r^2), \label{Rmatrix}
\end{equation}
satisfying the quantum Yang-Baxter 
equation (QYBE) which is said to be the universal quantum $R$--matrix
associated to the classical $r$--matrix.
The relevance of the previous procedure is that we can get many
``concrete'' solutions of the QYBE by taking different representations
of the universal object $\cal R$. Thus, if we take a finite
dimensional  representation $\rho$ of ${\fra g}$ in the algebra of
$n\times n$ complex matrices $M(n,\C)$, we have
$$
  R= (\rho \otimes \rho)({\cal R}) 
$$
which satisfies the QYBE
\begin{equation}
  R_{12} R_{13} R_{23} = R_{23} R_{13} R_{12},\label{qybe}
\end{equation}
and the unitary condition
\begin{equation}
    R R^\sigma = 1,\label{unicond}
\end{equation}
where $R^\sigma= (\rho \otimes \rho)(\sigma ({\cal R}))$.
We see that this quantization procedure leads in a natural way to 
QYBE.

Finally, let us consider again the matrix representation
$\rho:{\fra g}\longrightarrow M(n,\C)$, consequently the group $G$
is  realized as subgroup of $GL(n,\C)$. Let
$T=(t_{ij})_{i,j=1}^n$ be the matrix of coordinate functions on $G$
$$
   t_{ij}(g) = g_{ij}, \quad g\in G.
$$

Left and right actions of ${\fra g}$ on matrix coordinates on $G$ are
easily described using (\ref{vecfields}) by 
$$
\begin{array}{l}
  (X_L t_{ij})(g)= (gX)_{ij}= t_{ik}(g) X_{kj},  \\
  (X_R t_{ij})(g)= (Xg)_{ij}=  X_{ik} t_{kj}(g),  
\end{array}
\qquad \forall X \in {\fra g}.
$$

On the other hand, let $\hat{F}$ be the image of $F$ by
the representation $\rho$  (i.e. $\hat{F}= (\rho \otimes 
\rho)(F)$) and defining $T_1= T \otimes 1$ and $T_2= 1 \otimes T$,
the $*$--product between matrix coordinates of $G$ elements can be
expressed in an elegant manner by
$$
T_1 *_h T_2= \hat{F}^{-1} T \otimes T \hat{F},
$$
applying the flip operator to both sides of this expression we get
$$
  T_2 *_h T_1= \sigma (\hat{F}^{-1}) T 
\otimes T \sigma(\hat{F}).
$$
Taking into account expression (\ref{Rmatrix}) we obtain the relation
$$
   R T_1 *_h T_2 = T_2 *_h T_1 R,    
$$
which is the well-known formula that gives the commutation relations
between the matrix coordinate functions of $G$ defining the quantum
group $Fun_h(G)$. 


\subsect{Quantization of PL groups with MCYBE $r$--matrix }

In this case the procedure is similar to the previous one in the sense 
that we again define the $*$--product by expression (\ref{starprod}).
However, here the  difference is that condition  (\ref{conditionsF}b)
over $F$ (where $F=1-(h/2)r +o(h^2) \in U{\fra g}^{\otimes 2}[[h]]$
such that $(\epsilon \otimes {\mbox id})F= ({\mbox id} \otimes
\epsilon) F= 1$), which assures associativity, is now relaxed and
substituted  by the  more general \cite{Ta90}: 
$$
(F\otimes {\mbox id}) (\Delta_0 \otimes {\mbox id}) F= \alpha
( {\mbox id} \otimes F) ( {\mbox id}  \otimes \Delta_0) F, 
\quad \alpha \in U{\fra g}^{\otimes 3}[[h]] , 
$$
where a formal power series in $h$ with coefficients in 
$U{\fra g}^{\otimes 3}$ has been 
introduced.

The associativity of $*_h$ is assured if $\alpha$ is $G$-invariant,
i.e.,  
\begin{equation}
{\rm ad}_{{\fra g}}^{\otimes 3} \alpha=
[1\otimes 1 \otimes X +1\otimes X \otimes 1 + X\otimes 1 \otimes 1,  
\alpha]=0, \qquad \forall X \in {\fra g}.\label{ginvariance}
\end{equation}
Drinfel'd has proved the existence of  
$\alpha  \in U{\fra g}^{\otimes 3}[[h]]$ verifying
(\ref{ginvariance}) and other additional conditions that we do not
display here (for more details see \cite{Ta90}).

Also it is possible to construct an object by
$$
{\cal R}=\sigma(F^{-1}) e^{ht}F, 
$$
where t is an ${\rm ad}_{\fra g}^{\otimes 2}$--invariant 
symmetric element of ${\fra g} \otimes {\fra g}$ defined by 
$[[r,r]]= -[t_{13}, t_{23}]$. The matrix 
$$
R=( \rho\otimes\rho){\cal R}, 
$$
defined in terms of a matrix representation $\rho$ of ${\fra g}$
on $M(n,\C)$ satisfies the QYBE (\ref{qybe}), but it does not verify the
unitary condition (\ref{unicond}).

As in the previous case, if we consider the natural representation
$\rho$ of $\fra g$ on $M(n,\C)$, and the matrix representation of the
group elements, $T$, then the $*$--product of the matrix coordinate
functions of $G$ are given by 
\begin{equation}
T_1*_h T_2=\hat F^{-1}T\otimes T\hat F. \label{tt}
\end{equation}
On the other hand, 
$T_2*_h T_1=\sigma(\hat F^{-1})T\otimes T\sigma(\hat F)$, and from this
expression and (\ref{ttr}) one gets once more 
\begin{equation}
{R}T_1*_h T_2=T_2*_hT_1 {R}.\label{ttr}
\end{equation}

\subsect{Example 3: Quantization of the PL group $SL(2)$}

To illustrate all the above techniques of quantization of PL groups,
we present the  quantization of the PL group $SL(2)$.

A basis for the Lie algebra ${\fra sl}(2)$ is given
by three elements  $X_\pm$ and $H$ with commuting relations
$$
  [H, X_\pm]= \pm 2 X_\pm,  \qquad [X_+, X_-]= H.
$$
A classical $r$--matrix satisfying MCYBE for this algebra is
$$
   r=2 X_+ \wedge X_-= 
   X_+ \otimes X_- - X_- \otimes X_+ \in \Lambda^2 {\fra sl}(2).
$$
The associated Sklyanin bracket is
$$
\{f, g\}= X_+^R f X_-^R g-  X_-^R f X_+^R g
      - X_+^L f X_-^L g-  X_-^L f X_+^L g.
$$
A $2\times 2$ matrix representation of ${\fra sl}(2)$ is given by
$$
\rho(X_+)= \left(\begin{array}{cc} 0 & 1 \\ 0 & 0 \end{array} \right), 
\quad
\rho(X_-)= \left(\begin{array}{cc} 0 & 0 \\ 1 & 0 \end{array} \right), 
\quad
\rho(H)= \left(\begin{array}{cc} 0 & 1 \\ -1 & 0 \end{array} \right). 
$$
The $r$--matrix in this representation takes the explicit form 
$$
\hat{r}= \left( \begin{array}{cccc}
 0 &  0 & 0 & 0 \\ 0 & 0 & 1 & 0 \\ 
 0 & -1 & 0 & 0 \\ 0 & 0 & 0 & 0 \end{array} \right) .
$$
The elements of $SL(2)$ written in matrix coordinates are 
$$
T= \left(\begin{array}{cc} a & b \\ c & d \end{array} \right), 
$$
with the condition ${\mbox det} \, T= ad- bc = 1$.
  
The Poisson brackets for the group matrix coordinates can  be
directly computed by means of
$$
 \{T \stackrel{\otimes}{,}T\}= [\hat{r}, T \otimes T],
$$
which yields
$$
\begin{array}{ccc}
\{a, b\}= ab, & \quad \{a, c\}= ac, & \quad \{a,d\}= 2bc, \\
\{b, c\}= 0, & \quad \{b, d\}=bd, & \quad \{c,d\}= cd. \\
\end{array}
$$
Note that these relations  fix the Poisson brackets for all pairs of
functions in  ${\cal C}^\infty(SL(2))$ if we look at them as polynomials
in the variables  $a,b,c,d$. 

The quantization of $SL(2)$ is performed by means of 
$$
 \hat{F}= e^{-\frac{h}{2}\sigma} \left( \begin{array}{cccc}
    \sqrt{q} & 0 & 0 & 0 \\ 0 & u^{-1} & 0 & 0 \\
     0 & v & u & 0 \\ 0 & 0 & 0 & \sqrt{q} \end{array} \right), 
$$
where $q=e^h$, $u=\sqrt{\frac{2}{q+q^{-1}} }$ and 
$v=\frac{q-q^{-1}}{\sqrt{2(q+q^{-1})}}$. 
The corresponding representation for the $R$--matrix is 
$$
{R}_q= \sqrt{q} \sigma(\hat{F}^{-1}) e^{(\sigma-\frac{1}{2} I)h} 
   \hat{F}= \left( \begin{array}{cccc}
    q & 0 & 0 & 0 \\ 0 & 1 & 0 & 0 \\ 0 & q-q^{-1} & 1 & 0 \\
    0 & 0 & 0 & q \end{array} \right).
$$

A straightforward calculation using (\ref{tt}) gives the
$*$--product for coordinates $a,b,c,d$. And from 
(\ref{ttr})  one gets the usual 
relations defining the quantum group $SL_q(2)$:
$$
\begin{array}{ccc}
 a *_h b = q b *_h a,  & a *_h c = q c *_h a,  &  a *_h d -  d *_h a= 
q-q^{-1}) b *_h c, \\ 
 b *_h c = c *_h b, & b *_h d = q d *_h b,  &  c *_h d = q d *_h c.  
\end{array}
$$
 
\section*{Acknowledgments}  This work has been partially supported by
DGES of the  Ministerio de Educaci¢n y Cultura of Spain under Project
PB95-0719,  and the Junta de Castilla y Le¢n (Spain).


\end{document}